\documentclass[fleqn,usenatbib]{mn2e}
\usepackage{epsfig}
\usepackage{amsmath}
\usepackage[dvips]{color}

\begin{document}

\title[Cluster Magnification]{Cluster Magnification \& the Mass-Richness Relation \\ in CFHTLenS}

\author[J. Ford et al.]
{Jes Ford,$^1$ 
Hendrik Hildebrandt,$^{2,1}$ 
Ludovic Van Waerbeke,$^1$ 
Thomas Erben,$^{2}$ \newauthor 
Clotilde Laigle,$^{3,4,1}$
Martha Milkeraitis,$^{1}$
Christopher B. Morrison$^{2}$ \\
$^1$Department of Physics and Astronomy, University of British Columbia, 6224 Agricultural Road, Vancouver, B.C. V6T 1Z1, Canada \\
$^2$Argelander-Institut f\"ur Astronomie, Auf dem H\"ugel 71, 53121 Bonn, Germany \\
$^3$Institut d\'Astrophysique de Paris \& UPMC (UMR7095), 98 bis boulevard Arago, 75014 Paris, France \\
$^4$Ecole Polytechnique, 91128 Palaiseau Cedex, France
}

\maketitle

\setcounter{section}{0}
\setcounter{subsection}{0}
\setcounter{subsubsection}{0}

\begin{abstract}
Gravitational lensing magnification is measured with a significance of 9.7$\sigma$ on a large sample of galaxy clusters in the Canada-France-Hawaii Telescope Lensing Survey (CFHTLenS). This survey covers $\sim$154 deg$^2$ and contains over 18,000 cluster candidates at redshifts $0.2 \leq z \leq 0.9$, detected using the 3D-Matched Filter cluster-finder of \citet{Milkeraitis10}. We fit composite-NFW models to the ensemble, accounting for cluster miscentering, source-lens redshift overlap, as well as nearby structure (the 2-halo term), and recover mass estimates of the cluster dark matter halos in range of $\sim10^{13} M_\odot$ to $2\times10^{14} M_\odot$. Cluster richness is measured for the entire sample, and we bin the clusters according to both richness and redshift. A mass-richness relation $M_{200} = M_0 (N_{200} / 20)^\beta$ is fit to the measurements. For two different cluster miscentering models we find consistent results for the normalization and slope,  $M_0 = (2.3 \pm 0.2) \times 10^{13} M_\odot$, $\beta = 1.4 \pm 0.1$ and $M_0 = (2.2 \pm 0.2) \times 10^{13} M_\odot$, $\beta = 1.5 \pm 0.1$. We find that accounting for the full redshift distribution of lenses and sources is important, since any overlap can have an impact on mass estimates inferred from flux magnification.
\end{abstract}


\section{Introduction}
Clusters of galaxies are the most massive gravitationally bound structures in the Universe today. As such they can be useful cosmological probes, as well as laboratories for all kinds of interesting physics including galaxy evolution, star formation rates, and interactions of the intergalactic medium. There are several methods commonly used to estimate cluster masses (e.g., mass-to-light ratios, Xray luminosities, the Sunyaev-Zeldovich effect), but among them gravitational lensing is unique in being sensitive to all mass along the line of sight, irrespective of its type or dynamical state \citep{BS01}.

There are multiple ways to measure the signature of gravitational lensing, and each has its own specific advantages and limitations. Observation of strong lensing arcs and multiple images is extremely useful for studying the innermost regions of clusters, and getting precise mass estimates, but can only be applied to very massive objects which are observationally limited in number. On the other hand, weak lensing shear, which measures slight deformations in background galaxy shapes, can be applied across a much wider range of lens masses. Shear studies have been used with much success to map large scale mass distributions in the nearby universe \citep{Waerbeke13, Massey07}. However, because they rely on precise shape measurements, shear faces the practical limitation of an inability to sufficiently resolve sources for lenses more distant than a redshift of about one \citep{LHJM10}.

A third approach to measuring gravitational lensing is through the magnification of background sources, observable either through source size and flux variations \citep{Schmidt12,Huff14}, or the resultant modification of source number densities \citep{Ford12, Morrison12, Hildebrandt13, Hildebrandt11, Hildebrandt09b, Scranton05}. Magnification has been recently measured using quasar variability as well \citep{Bauer11}. Although relative to the shear, magnification will tend to have a lower signal-to-noise for typical low-redshift lenses, the requirement for source resolution is completely removed. This makes magnification competitive for higher redshift lenses, and especially for ground based surveys where atmospheric seeing has a strong influence on image quality.

In this work we adopt the number density approach, known as flux magnification, using Lyman-break galaxies (LBGs) for the lensed background sources. The observed number density of LBGs is altered by the presence of foreground structure, due to the apparent stretching of sky solid angle, and the consequential amplification of source flux. Because of the variation in slope of the LBG luminosity function, magnification can either increase or decrease the number densities of LBGs, depending on their intrinsic magnitudes. By stacking many clusters we can overcome the predominant source of noise - physical source clustering \citep{Hildebrandt11}. 

In Section \ref{data} we describe the cluster and background galaxy samples. Section \ref{method} lays out the methodology for the measurement and modeling of the magnification signal. We discuss our results in Section \ref{results} and conclude in Section \ref{conc}. Throughout this paper we give all distances in physical units, and use a standard $\Lambda$CDM cosmology with $H_0 =$ 70 km s$^{-1}$ Mpc$^{-1}$, $\Omega_M = 0.3$, and $\Omega_{\Lambda} = 1 - \Omega_M = 0.7$.


\section{Data}
\label{data}
For the magnification results presented in this paper, we are fortunate to work with a very large sample of galaxy cluster candidates and background galaxies in the Canada-France-Hawaii-Telescope Lensing Survey (CFHTLenS\footnote[1]{www.cfhtlens.org; \\Data products available at http://www.cadc-ccda.hia-iha.nrc-cnrc.gc.ca/\-community/\-CFHTLens/\-query.html}; \citet{Erben13}; \citet{Hildebrandt12}). CFHTLenS is based on the Wide portion of the Canada-France-Hawaii Telescope Legacy Survey, with deep 5 band photometry. The survey is composed of four separate fields, in turn divided into 171 individual pointings, covering a total of 154 deg$^2$.

\subsection{3D-MF Galaxy Clusters}
\label{clusters}
The 3D-Matched-Filter (3D-MF) cluster finding algorithm of \citet{Milkeraitis10}, essentially creates likelihood maps of the sky (in discrete redshift bins) and searches for peaks of significance above the galaxy background. The likelihood is estimated assuming that clusters follow a radial Hubble profile as well as a Schechter luminosity function. A significance peak of \textgreater 3.5$\sigma$ is considered a cluster detection, since this reduces below 1\% the probability of Gaussian random noise fluctuations mimicking a true cluster. The reader is referred to \citet{Milkeraitis10} for the details of the 3D-MF algorithm; here we discuss only the essential points relevant to our purposes. 

The radial component of the 3D-MF likelihood employs a cutoff radius of 1 Mpc, which was chosen to roughly correspond to the radius $r_{200}$ of an $M_{200} \sim 3 \times 10^{13} M_{\odot}$ cluster. \citet{Milkeraitis10} motivates this choice by the desire to optimally search for relatively high mass clusters, but notes that this radius will be less ideal for low mass clusters. One should expect that random galaxy interlopers may contaminate the estimation of significance for likelihood peaks corresponding to lower mass clusters. This may be a key factor in explaining the wide range of cluster significances conferred upon low mass clusters from simulations, while high mass simulated clusters were assigned significances that correlated strongly with mass \citep[see Figure 10 in][]{Milkeraitis10}. Because peak significance may therefore not be an ideal mass proxy to use for the full cluster ensemble, in this work we rely upon a measure of the cluster richness, which is discussed in Section \ref{rich} below. Using cluster richness has the added benefit that the mass-richness relation can be measured and used as a scaling relation.

Using the 3D-MF method, a total of 18,036 galaxy cluster candidates (hereafter clusters) have been detected in CFHTLenS, at a significance of \textgreater 3.5$\sigma$ above the background. In contrast to previous cluster magnification studies, which have been limited by small number statistics, this huge sample of clusters allows us to pursue multiple avenues of investigation. In particular, we bin the clusters according to both richness and redshift, to recover trends in physical characteristics such as the mass-richness relation, and also investigate halo miscentering as a function of these parameters.

\subsubsection{Cluster Centers}
\label{centering}
Due to the nature of the method, the defined centers of the 3D-MF clusters, which are located at peaks in the likelihood map, do not necesssarily coincide with member galaxies. Hence the defined centers are notably different from many other cluster finders, which commonly choose the brightest cluster galaxy (BCG), the peak in X-ray emission, some type of (possibly luminosity-weighted) average of galaxy positions, or a combination of these, as a measure of the center of a dark matter halo.

\begin{figure}
\begin{center}
\includegraphics[scale=0.4]{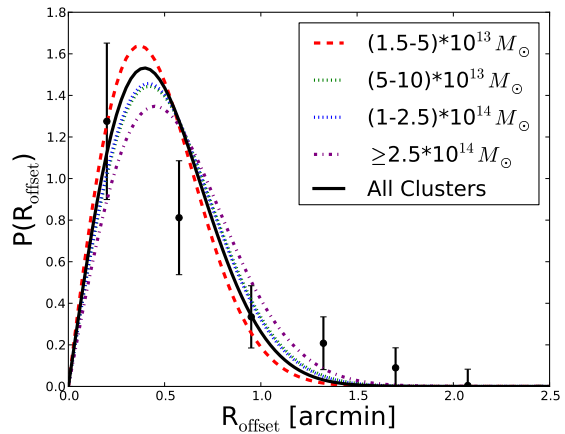}
\caption{Modeled Gaussian distribution of radial offsets between defined 3DMF centers and simulated cluster centers. The black points and solid curve is the combined data and best fit for all CFHTLenS clusters combined. The colored curves show the best fit Gaussians for separate mass bins, and colors match the empirical offsets measured and presented in Figure 13 of \citet{Milkeraitis10}. As each of these colored curve fits is consistent with the solid black curve for the entire combined sample, we choose to use this single Gaussian distribution to model miscentering for all clusters.}
\label{gauss}
\end{center}
\end{figure}

\begin{table}
  \begin{center}
    \caption{Best Fit Gaussian Distributions for the Cluster Miscentering in Figure \ref{gauss}.}
    \begin{tabular}{llll}
      \hline
      Mass Range [$M_{\odot}$] & Color & Best fit $\sigma_{\mathrm{offset}}$ [arcmin] & $\chi^2_{red}$ \\ \hline
      (1.5-5)$\times10^{13}$ & red    & 0.37$\pm$0.06 & 1.8  \\
      (5-10)$\times10^{13}$  & green  & 0.42$\pm$0.06 & 1.4  \\
      (1-2.5)$\times10^{14}$ & blue   & 0.42$\pm$0.06 & 1.2  \\ 
      $\geq$2.5$\times10^{14}$   & purple & 0.45$\pm$0.06 & 1.4  \\ \hline
      $\geq$1.5$\times10^{13}$  & black  & 0.40$\pm$0.06 & 1.1  \\
      \hline
    \end{tabular}
  \end{center}
\end{table}

The choice of cluster center is always ambiguous, both observationally and in simulations. One wants to know the center of the dark matter distribution, as the point around which to measure a radially-dependent signal. Obviously the dark matter cannot be directly seen, so an observable such as galaxies or X-ray emission must be used (see \citet{George12} for an excellent review and analysis of cluster centroiding). The chosen center of the cluster can be wrong for several reasons. The observable chosen (e.g. the BCG) may simply be offset from the true center of the dark matter potential. Misidentification of the BCG can be a significant problem for this particular example as well \citep{Johnston07}. 

Perhaps a more interesting source for miscentering comes from the fact that clusters halos are not perfectly spherical, and exhibit substructure and irregularities caused by their own unique mass assembly histories. Especially for very massive halos, which have formed more recently and in many cases are still undergoing mergers and have yet to virialize, we really should not expect a clear center to exist. Following visual inspection, \citet{Mandelbaum08} chose to exclude the most massive clusters from their weak lensing analysis for this very reason. Instead of throwing away the highest mass halos in our sample, we include them in this study, but take care to account for possible miscentering effects. 

\citet{Milkeraitis10} tested for centroid offsets in 3D-MF by running the cluster-finder on simulations and comparing detected cluster centers to known centers. The simulations used were the mock catalogs of \citet{KW07}, which were created from a semi-analytic galaxy catalog \citep{DeLucia07} derived from the Millenium Simulation \citep{Springel05}. Figure 13 of that work shows the number of clusters detected as a function of distance from true cluster center. Because 3D-MF was optimized to produce cluster catalogs that are as complete as possible (in contrast to, e.g. \citet{Gillis11}, which is designed to maximize purity), the trade-off is the presence of some contamination with false-detections, especially at the low mass end.

We use the numbers of clusters at each offset, and the contamination from \citet{Milkeraitis10}, to estimate the probability of radial offsets $P(R_{\mathrm{offset}})$. We fit the result for each mass bin with a 2-dimensional Gaussian distribution:
\begin{equation}
P(R_{\mathrm{offset}}) = \frac{R_{\mathrm{offset}}}{\sigma_{\mathrm{offset}}^2} \exp \left[ -\frac{1}{2}\left( \frac{R_{\mathrm{offset}}}{\sigma_{\mathrm{offset}}} \right)^2 \right].
\label{PofRc}
\end{equation}
This resulting curves are presented in Figure \ref{gauss} (colors are selected to match those in Figure 13 of \citet{Milkeraitis10}), and for clarity we only show the data points for the bin that combines all clusters. We find consistent fits for the separate mass bins, which we list in Table 1, and therefore use the combined distribution (black curve) to model the effects of miscentering in our measurements.

\subsubsection{Cluster Richness}
\label{rich}
We define the richness parameter $N_{200}$ in this work to be the number of galaxies within a radius $r_{200}$, and redshift $\Delta z$, of a cluster candidate center (both points discussed below). Member galaxies are also required to be brighter than $i$-band absolute magnitude -19.35. This cut-off is chosen to correspond to the limiting {\it apparent} magnitude of CFHTLenS ($i \sim 24$) at the highest redshift clusters that we probe, $z \sim 1$. So at the expense of removing many galaxies from the richness count, we hope to largely avoid the effect of incompleteness on the number of galaxies per cluster. Then clusters of the same intrinsic richness at high and low redshift should have comparable observed $N_{200}$, within the expected scatter of the mass-richness relation.

For the line-of-sight dimension, we require galaxies to fall within $\Delta z < 0.08(1+z)$ of the cluster redshift. This $\Delta z$ is the 2$\sigma$ scatter of photometric redshifts in the CFHTLenS catalog, chosen so that we reduce the probability of galaxies in a cluster being missed due to errors in their photo-$z$ estimation. Of course this comes at the expense of counting galaxies within a quite broad line-of-sight extent, especially for the higher redshift clusters. This effect should cancel out though, since we also use the same $\Delta z$ range in calculating the galaxy background density, which is subtracted to yield $N_{200}$ as an overdensity count of cluster galaxy members.

\begin{figure}
\begin{center}
\vspace{0.5 cm}
\includegraphics[scale=0.4]{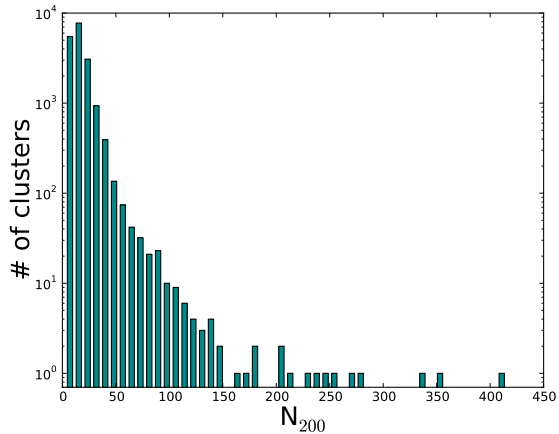}
\caption{Distribution of richness ($N_{200}$) values for clusters in this study.}
\label{hist}
\end{center}
\end{figure}

In the plane of the sky, galaxies must lie within a projected radius $r_{200}$ of the cluster center (defined above). $r_{200}$ is defined as the radius within which the average density is 200 times the critical energy density of the universe, $\rho_{crit}(z)$, evaluated at the redshift of the cluster. However, since $r_{200}$ itself is unknown, we require some kind of assumption about radius or mass in order to proceded with the galaxy counting. There is no unique way to do this. We begin by making an initial approximation of the masses using a best fit power-law relation between mass and cluster peak significance, for 3D-MF clusters \citep{MMthesis11}:
\begin{equation}
\mathrm{log}(M_{200}) = 0.124 \sigma + \mathrm{log}(10^{14} M_{\odot}) - 1.507.
\end{equation}

As discussed in \ref{clusters}, 3D-MF tests on simulations suggested that peak significance was a good mass proxy for high, but not low, mass clusters. In light of this, we merely employ the above relation as a starting point for calculating the radii from mass,
\begin{equation}
r_{200} = \left[ \frac{3 M_{200}}{4 \pi (200) \rho_{crit}(z)} \right] ^ {1/3}.
\end{equation}
These $r_{200}$ estimates are then used for counting galaxies for cluster richness. Richness $N_{200}$ is the variable of choice used as a mass proxy for binning the magnification measurement. The distribution of these richness values is shown in Fig. 2.


\subsection{Sources}
We use Lyman-break galaxies (LBGs) as the magnified background sources. LBGs are high-redshift star-forming galaxies \citep{Steidel98}, that have been succesfully employed in past magnification studies \citep[see][]{Hildebrandt09b, Hildebrandt11, Morrison12, Ford12} due to the fact that their redshift distributions and luminosity functions are reasonably well understood. Knowledge of the intrinsic source luminosity function allows for an interpretation of the magnification signal, which depends sensitively on the slope of the number counts as a function of magnitude. In addition, the high-redshift nature of LBGs is important to reduce redshift overlap between lenses and sources. Any source galaxies in the redshift range of the cluster lenses will contaminate the lensing-induced cross-correlation signal, with correlations due to physical clustering.

The LBG sample is selected with the color selection criteria of \citet{Hildebrandt09a} (see Sect.~3.2 of that paper). It is composed of 122,144 $u$-dropouts with 23 \textless $r \leq$ 24.5, located at redshift $\sim$3.1. We choose this magnitude range to avoid as much potential low-redshift contamination as possible. See Section \ref{contam} for our modeling of the residual contamination. The detailed properties of this LBG population will be described in a forthcoming paper (Hildebrandt et al. in prep.).


\section{Magnification Methodology}
\label{method}

\subsection{The Measurement}

The magnification factor, $\mu$, of a gravitational lens can be expressed in terms of the change from intrinsic ($n_0$) to observed ($n$) differential number counts of background sources:
\begin{equation}
n(m)dm = \mu^{\alpha-1} n_0(m)dm 
\end{equation}
\citep{Narayan89}. Here $m$ is apparent magnitude, and $\alpha \equiv \alpha(m)$ is proportional to the logarithmic slope of the source luminosity function. Depending on the luminosity function's slope in a given magnitude bin, it is possible to observe either an increase or a decrease in source number counts, as demonstrated in Figure 2 of \citet{Ford12}. 

In practice the magnification signal is easily measured using the optimally-weighted cross-correlation function of \citet{Menard03}:
\begin{equation}
\mathrm{w}_{\mathrm{opt}}(R)=\frac{S^{\alpha-1} L - S^{\alpha-1} R - \langle \alpha-1 \rangle LR}{RR} + \langle \alpha-1 \rangle.
\end{equation}
In this expression, the terms are normalized pair counts in radial bins, where $L$ stands for the lenses, and $S^{\alpha-1}$ are the optimally-weighted sources. $R$ represents objects from a random catalog more than ten times the size of the source catalog, with the same masks applied.

In order to determine the optimal weight factor $\alpha-1$, for both the measurement and the interpretation, we require knowledge of the source luminosity function. As done in \citet{Ford12} we determine the LBG luminosity function slope from the Schechter Function \citep{Schechter76}, giving
\begin{equation}
\alpha = 2.5 \frac{\mathrm{d}}{\mathrm{d}m}\log n_0(m) = 10^{0.4(M^\ast-M)}-\alpha_{\mathrm{LF}}-1,
\end{equation}
and rely on externally measured luminosity functions for the characteristic magnitude $M^\ast$ and faint end slope $\alpha_{\mathrm{LF}}$. We use the LBG luminosity function of \citet{vanderBurg10}, measured using much deeper data from the CFHTLS Deep fields. For $u$-dropouts $M^\ast$ is -20.84 and $\alpha_{\mathrm{LF}}$ is -1.6. Thus every source galaxy is assigned a weight factor of $\alpha-1$ according to its absolute magnitude $M$.

The magnification signal, $\mathrm{w}_{\mathrm{opt}}(R)$, is measured in logarithmic radial bins of physical range 0.09 -- 4 Mpc (in contrast to angle), so that we can stack clusters at different redshifts without mixing very different physical scales. Each cluster's signal is measured separately before stacking the measured $\mathrm{w}_{\mathrm{opt}}(R)$, and full covariance matrices are estimated from the different measurements.

\subsection{The Modeling}
\label{model}

The magnification is a function of the halo masses, and to first order it is proportional to the convergence $\kappa$. In this work, however, we will use the full expression for $\mu$ to account for any deviations from weak lensing in the inner regions of the clusters:
\begin{equation}
\mu = \frac{1}{(1-\kappa)^2 - \left|\gamma\right|^2}
\end{equation}
\citep{BS01}.

We assume a spherical Navarro-Frenk-White (NFW) model \citep{nfw97} for the dark matter halos, along with the mass-concentration relation of \citet{Prada12}. The convergence is modeled as the sum of three terms,
\begin{equation}
\kappa = \left[p_{\mathrm{cc}}\Sigma_{\mathrm{NFW}} +(1-p_{\mathrm{cc}})\Sigma_{\mathrm{NFW}}^{smoothed} + \Sigma_{\mathrm{2halo}} \right]/\Sigma_{\mathrm{crit}},
\label{modelEQ}
\end{equation}
where $p_{\mathrm{cc}}$ is the fraction of clusters correctly centered (i.e. with $R_{\mathrm{offset}}=0$), and $\Sigma_{\mathrm{crit}}(z)$ is the critical surface mass density at the lens redshift. The expression for the shear, $\gamma$, is identical with $\kappa \rightarrow \gamma$, and $\Sigma \rightarrow \Delta\Sigma$. Note that the first term in Equation \ref{modelEQ} is equivalent to adding a delta function to the miscentered distribution of Figure \ref{gauss}, to represent clusters with perfectly-identified centers. As discussed in Section \ref{results}, the fits do not give strong preference to miscentering in the measurement, but in future work (in particular with weak lensing shear) it will be useful to constrain the degree of miscentering using the data, instead of relying solely on simulations.

We assume both lenses and sources are located at known discrete redshifts. This is $z \sim$ 3.1 for the LBGs. Since they are at very high redshift the effect of any small offsets from this has negligible effect on the angular diameter distance, the relevant distance measure for lensing. The clusters, on the other hand, have redshift uncertainties of 0.05 (due to the shifting redshift slices employed by 3D-MF \citep{Milkeraitis10}). This translates into an uncertainty on the mass estimates ranging from less than a percent up to $\sim$17\% (depending on cluster $z$), and is included in the reported mass estimates.

$\Sigma_{\mathrm{NFW}}$ is the standard surface mass density for a perfectly centered NFW halo, calculated using expressions for $\kappa$ (and $\gamma$) in \citet{Wright00}. $\Sigma_{\mathrm{NFW}}^{smoothed}$ on the other hand, is the expected surface mass density measured for a miscentered NFW halo:

\begin{equation}
\Sigma_{\mathrm{NFW}}^{smoothed}(R)  = \int_0^\infty \Sigma_{\mathrm{NFW}}(R \vert R_{\mathrm{offset}}) P(R_{\mathrm{offset}}) \mathrm{d}R_{\mathrm{offset}}.
\end{equation}
The distribution of offsets $P(R_{\mathrm{offset}})$ is given by Equation \ref{PofRc}, and the other factor in the integrand is
\begin{equation}
\Sigma_{\mathrm{NFW}}(R \vert R_{\mathrm{offset}}) = \frac{1}{2\pi} \int_0^{2\pi} \Sigma_{\mathrm{NFW}}(R') \mathrm{d}\theta,
\end{equation}
where $R' = \sqrt{R^2 + R_{\mathrm{offset}}^2 + 2RR_{\mathrm{offset}}\cos\theta}$ \citep{Yang06}.

The 2-halo term $\Sigma_{\mathrm{2halo}}$ accounts for the fact that the halos we study do not live in isolation, but are clustered as all matter in the universe is. We account for neighboring halos following the prescription of \citet{Johnston07}:

\begin{equation}
\Sigma_{\mathrm{2halo}}(R,z) = b_{l}(M_{200},z) \Omega_M \sigma_8^2 D(z)^2 \Sigma_l(R,z)
\end{equation}
\begin{equation}
\Sigma_l(R,z) = (1+z)^3 \rho_{crit,0} \int_{-\infty}^\infty \xi\left( (1+z)\sqrt{R^2 + y^2} \right) \mathrm{d}y
\end{equation}
\begin{equation}
\xi(r) = \frac{1}{2\pi^2} \int_0^\infty k^2 P(k) \frac{\sin{kr}}{kr} \mathrm{d}k
\end{equation}
Here small $r$ is comoving distance, $D(z)$ is the growth factor, $P(k)$ is the linear matter power spectrum, and $\sigma_8$ is the amplitude of the power spectrum on scales of 8 $h^{-1}$Mpc. For the lens bias factor $b_{l}(M_{200},z)$ we use Equation 5 of \citet{Seljak04}.

\subsubsection{Composite-Halo Fits}
\label{multihalo}
The part of the optimal correlation function which is caused by gravitation lensing is related to the magnification contrast $\delta\mu \equiv \mu-1$ through
\begin{equation}
\label{wmodel}
\mathrm{w}_{\mathrm{lensing}}(R) = \frac{1}{N_{lens}} \sum_{i=1}^{N_{lens}} \langle(\alpha-1)^2\rangle_i \delta\mu(R,M_{200})_i.
\end{equation}
Here the sum is over the number of lenses in a given stacked measurement, and $\langle(\alpha-1)^2\rangle_i$ refers to the average of the weight factor squared in the pointing of a given cluster.

We perform composite-halo fits using the above prescription, in which we allow for the fact that the clusters in a given measurement have a range of masses and redshifts. We do not fit a single average mass. Instead, we calculate $\delta\mu(R,M_{200})_i$ for each individual cluster using a scaling relation between mass and richness,
\begin{equation}
\label{mr}
M_{200} = M_0 \left( \frac{N_{200}}{20} \right)^\beta.
\end{equation}
The fit parameters are the normalization, $M_0$, and (log-) slope, $\log\beta$, of the assumed power-law relation. From this we calculate the optimal correlation $\mathrm{w}_{\mathrm{opt}}(R)$ according to Equation \ref{wmodel}. The best-fit relation is determined by minimizing $\chi^2$, which is calculated using the full covariance matrix. We apply the correction factor from \citet{Hartlap07} to the inverse covariance matrix; this corrects for a known bias (related to the number of data sets and bins) which would otherwise lead to our error bars being too small.

\subsubsection{LBG Contamination}
\label{contam}
An important source of systematic error for magnification comes from low-redshift contamination of the sources, leading to physical clustering between the lens and source populations. The cross-correlation that results from contamination can easily overwhelm the measurement of magnification, making redshift overlap far more important for magnification than for shear. Past studies sought to minimize this effect, for example by checking for the negative cross-correlation that should exist between lenses and very faint sources with shallow number count \citep{Ford12, Hildebrandt09b}. Here we incorporate this clustering into the model, using a similar approach to \citet{Hildebrandt13}.

\begin{figure}
\begin{center}
\vspace{0.5cm}
\includegraphics[scale=0.5]{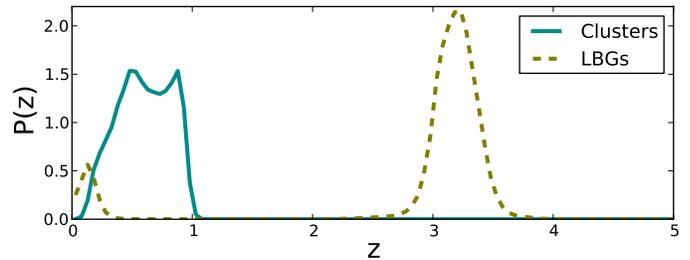}
\caption{Redshift probability distribution functions for the clusters and the LBG sources. Low-redshift contamination of the LBGs will lead to physical clustering correlations where overlap with the cluster redshifts occurs.}
\label{pofz}
\end{center}
\end{figure}

Figure \ref{pofz} shows the redshift probability distributions, $P(z)$, for the clusters and the LBGs. The LBG redshift distribution is based on the stacked posterior $P(z)$ put out by the BPZ redshift code \citep[for details on the CFHTLenS photo-$z$ see][]{Hildebrandt12}. Since the BPZ prior is only calibrated for a magnitude limited sample of galaxies we can not expect the stacked $P(z)$ to reflect the real redshift distribution of the color-selected LBGs. Hence we use the location and shape of the primary (high-$z$) and secondary (low-$z$) peaks but adjust their relative heights separately. This can be done with a cross-correlation technique similar to \citet{Newman08}. Details of this technique will be presented in Hildebrandt et al. (in prep.).

Despite our efforts to avoid contamination, there is obviously some redshift overlap with the clusters. We use the products of the lens and source $P(z)$ to define selection functions, and calculate the expected angular correlations using the code from \citet{Hamana04}. The weighted correlation function that we measure is the sum of the correlations due to lensing magnification and clustering contamination:
\begin{equation}
\mathrm{w}_{\mathrm{opt}}(R,z) = f_{\mathrm{lensing}}\mathrm{w}_{\mathrm{lensing}} + f_{\mathrm{clustering}}\mathrm{w}_{\mathrm{clustering}}.
\end{equation}
Note that $f_{\mathrm{lensing}}+f_{\mathrm{clustering}} \leq 1$, since some of the contaminants may be neither in the background and lensed, nor close enough in redshift to be clustered with the lenses.

The clustering contamination fraction $f_{\mathrm{clustering}}(z)$ for each cluster redshift is defined as the fraction of each source $P(z)$ that lies within 0.1 in redshift (twice the cluster redshift uncertainty). The part of the source $P(z)$ that lies at higher redshift than the lens is then the lensed fraction $f_{\mathrm{lensing}}(z)$, and the part at lower redshift (i.e. in {\it front} of the lens) has no contribution to the signal.

The factor $f_{\mathrm{clustering}}(z)$ itself is generally very small for the LBGs used in this work, only really non-negligible for cluster redshifts $z \sim$ 0.2-0.3, which can be seen in Figure \ref{pofz}. The more significant effect on the estimated masses is that $f_{\mathrm{lensing}}(z) \sim$0.9 across all redshift bins, because about 10\% of the sources are not really being lensed. We tested our results for robustness against uncertainties in the contamination fraction. When we vary the total low-$z$ contamination fraction by $\pm1\sigma$ ($\sim$4\%), the best fit cluster mass estimates remain within the stated error bars.

We explore three ways of determining $\mathrm{w}_{\mathrm{clustering}}$. Because of the weighting applied to LBGs in our measurement (which is optimal for the lensed sources, and should suppress contributions from redshift overlap), there will always be a prefactor of $\langle \alpha-1 \rangle$ in each estimation of clustering. The first method uses the dark matter angular auto-correlation, $\mathrm{w}_{\mathrm{dm}}$, and estimates of the galaxy and cluster bias to calculate:
\begin{equation}
\mathrm{w}_{\mathrm{clustering}}(R,z) = \langle \alpha-1 \rangle b_{l} b_{s} \mathrm{w}_{\mathrm{dm}}(R,z).
\label{model1}
\end{equation}
We set the bias factor for the galaxy contaminants $b_s$=1 for this analysis, which is reasonably consistent with the bias relation of \citet{Seljak04} that is employed for the cluster bias ($b_{l}$).

We also calculate both the 1- and 2-halo terms for NFW halos, w$_{\mathrm{1halo}}$ and w$_{\mathrm{2halo}}$ \citep[again using the code and methods described in][]{Hamana04}. Here the expression for physical clustering takes the form:
\begin{equation}
\mathrm{w}_{\mathrm{clustering}}(R,z) = \langle \alpha-1 \rangle b_{s} \left[\mathrm{w}_{\mathrm{1halo}}(R,z)+\mathrm{w}_{\mathrm{2halo}}(R,z)\right].
\end{equation}
This method requires knowledge of the occupation distribution of the low-$z$ galaxy contaminants in the cluster dark matter halos, which is not well determined. As a first approximation we use the simple power-law form described in \citet{Hamana04},
\begin{equation}
N_g(M) = \begin{cases} (M/M_1)^\alpha & \text{for $M > M_{\mathrm{min}}$} \\ 0 & \text{for $ M < M_{\mathrm{min}}$} \end{cases}.
\end{equation}
Since these parameters are unknown, we use the values for $M_1$ and $\alpha$ measured for galaxies in the SDSS \citep[see Table 3 of][]{Zehavi11}. We choose $M_{\mathrm{min}}$ to correspond to the minimum mass measured for cluster halos, and assume that halos above this mass always host a detected cluster. As a final check, we also ask what the clustering signal would be if every halo above $M_{\mathrm{min}}$ hosted both a cluster and a single low-$z$ galaxy contaminant (so that $N_g = 1$ for $M > M_{\mathrm{min}}$).

This final method yields the largest estimates of $\mathrm{w}_{\mathrm{clustering}}$, and therefore a smaller estimate of cluster masses. The former (using SDSS parameters) gives the highest mass estimates, and the simple biasing approach of Equation \ref{model1} yields intermediate results. We use the range of these results to estimate an uncertainty in mass estimates coming from lack of knowledge about the nature of the low-$z$ galaxies that contaminate our LBG source sample. This additional systematic error affects only the clusters at low redshift, where the source and lens $P(z)$ distributions overlap, and is reported on the mass estimates given in Table 3. All best fit mass values reported in the tables of this work are calculated using the contamination approach of Equation \ref{model1}, since this method relied on the fewest assumptions about the nature of the galaxy contaminants.

Accounting for redshift distributions in this particular source sample effectively means that cluster masses are {\it higher} than one would naively guess by fitting for only the magnification signal. However, note that in a case with more significant redshift overlap, so that $f_{\mathrm{clustering}}$ was large, the opposite statement would be true, and mass estimates that included the full $P(z)$ distribution would be smaller than than the naive magnification-only approach. These are important effects to consider, and future flux magnification studies should be careful to use full redshift distributions in modeling the measured signal.

\begin{figure*}
\begin{center}
\vspace{0.5 cm}
\includegraphics[scale=0.55]{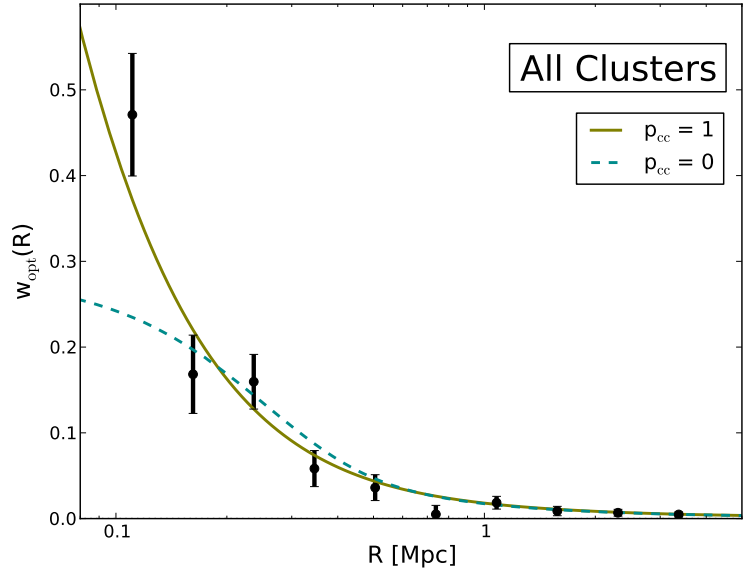}
\caption{Optimal cross-correlation signal measured for the entire stacked sample of 18,024 clusters. The model fits are both composite-NFW (see text for all terms in the fit). The solid line assumes the clusters are perfectly centered on the peak likelihood of the 3DMF cluster detection, while the dashed line includes the effects of cluster miscentering.}
\vspace{1.5 cm}
\label{w_all}
\end{center}
\end{figure*}


\section{Results}
\label{results}

Stacking the entire set of 18,036 clusters gives a total significance of 9.7$\sigma$ for the combined detection, shown in Figure \ref{w_all}. The perfectly centered model is a better fit to the overall measurement, with $\chi^2_{red}$ $\sim$ 1.2, while the miscentered model gives $\chi^2_{red}$ of 2.3. For both models, there are two free parameters ($M_0$ and $\log\beta$), leading to 8 degrees of freedom. To investigate miscentering and mass-richness scaling, we divide the clusters into six richness bins, and measure the optimal cross-correlation in each. 

\begin{figure*}
\begin{center}
\includegraphics[scale=0.7]{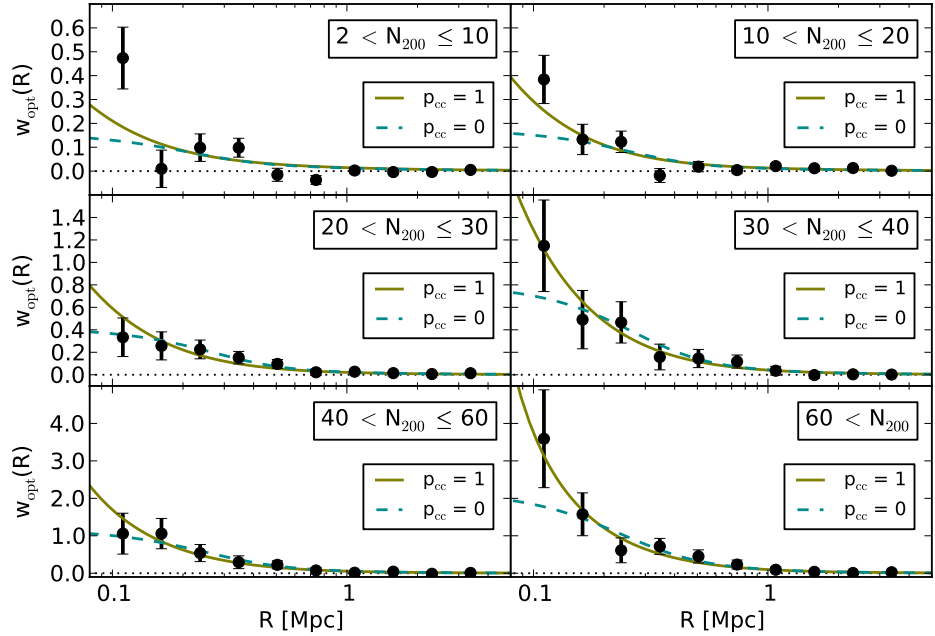}
\caption{Optimal cross-correlation signal measured for each $N_{200}$ (richness) bin. Two composite-NFW fits are shown. The solid curve assumes clusters are perfectly centered, while the dashed curve accounts for cluster miscentering, using the gaussian offset distribution modeled in Figure \ref{gauss} and discussed in Section \ref{model}.}
\label{binned}
\end{center}
\end{figure*}

We measure the characteristic signature of magnification in every richness bin with significances between 4.6 and 5.9$\sigma$. These results are shown in Figure \ref{binned}, where we try fitting both a perfectly centered model ($p_{\mathrm{cc}}=1$) and a model where every cluster is affected by centroid offsets ($p_{\mathrm{cc}}=0$). Details of the fits, including reduced $\chi^2$ and the average of the best fit mass values $\langle M_{200} \rangle$, are given in Table \ref{richtable}.

The lowest mass (richness) bin is not well fit by either model. Overall there is not a strong preference for either perfectly centered ($p_{\mathrm{cc}}=1$) or miscentered ($p_{\mathrm{cc}}=0$) clusters, and both are reasonably good fits. Generally, the miscentered model yields slightly higher masses for the clusters (though it is sensitive to the shape of the data), due to the Gaussian smoothing applied, which lowers the model amplitude in the innermost regions. However this is easily within the uncertainty on the mass estimates, so the results are in agreement.

\begin{table*}
  \begin{minipage}{16.cm}
    \label{richtable}
    \caption{Details of fits for richness-binned measurements in Figure \ref{binned}. We list the richness range selected, the number of clusters in that bin, the detection significance, the average richness of the bin, and the mass estimates and reduced $\chi^2$ for both the centered and miscentered models fit to the data. Note that the average mass given is not the value fit itself, but the average of all resulting masses fit using the composite-halo approach discussed in Section \ref{multihalo}.}
    \begin{tabular}{llllllll}
      \hline
      Richness & \# Clusters & Significance & $\langle N_{200} \rangle$ & p$_{cc}$=1: $\langle M_{200} \rangle$ & $\chi^2_{red}$ & p$_{cc}$=0: $\langle M_{200} \rangle$ & $\chi^2_{red}$ \\ \hline
      2 $<N_{200}$ & 18036 & 9.7$\sigma$ & 17 & (2.0$\pm$0.3)$\times10^{13} M_{\odot}$ & 1.2 & (1.8$\pm$0.3)$\times10^{13} M_{\odot}$ & 2.3  \\
      2 $<N_{200}<$ 10 & 4453 & 5.3$\sigma$ & 8 & (0.9$\pm$0.5)$\times10^{13} M_{\odot}$ & 3.0 & (0.7$\pm$0.4)$\times10^{13} M_{\odot}$ & 3.2  \\
      10 $<N_{200}<$ 20 & 9398 & 5.9$\sigma$ & 15 & (1.3$\pm$0.3)$\times10^{13} M_{\odot}$ & 1.6 & (1.0$\pm$0.3)$\times10^{13} M_{\odot}$ & 2.2  \\
      20 $<N_{200}<$ 30 & 2967 & 5.4$\sigma$ & 24 & (2.9$\pm$0.7)$\times10^{13} M_{\odot}$ & 0.7 & (3.3$\pm$0.8)$\times10^{13} M_{\odot}$ & 0.3  \\
      30 $<N_{200}<$ 40 & 695 & 5.0$\sigma$ & 35 & (7$\pm$2)$\times10^{13} M_{\odot}$ & 0.3 & (7$\pm$2)$\times10^{13} M_{\odot}$ & 0.5  \\
      40 $<N_{200}<$ 60 & 351 & 4.6$\sigma$ & 47 & (1.0$\pm$0.2)$\times10^{14} M_{\odot}$ & 0.4 & (1.1$\pm$0.2)$\times10^{14} M_{\odot}$ & 0.3  \\
      60 $<N_{200}$ & 172 & 5.5$\sigma$ & 99 & (2.0$\pm$0.4)$\times10^{14} M_{\odot}$ & 0.5 & (2.1$\pm$0.4)$\times10^{14} M_{\odot}$ & 0.6  \\
      \hline
    \end{tabular}
  \end{minipage}
\end{table*}

The issue of cluster miscentering is interesting in its own right as discussed in Section \ref{centering}. It is tempting to try and fit for the parameter $p_{\mathrm{cc}}$, describing the fraction that are actually correctly centered, or else for the miscentering Gaussian width $\sigma_{\mathrm{offset}}$, as done in \citet{Johnston07}. The issue here is a strong degeneracy between $p_{\mathrm{cc}}$, $\sigma_{\mathrm{offset}}$, and cluster concentration. Increasing the number of clusters that have offset centers produces essentially the same results as leaving $p_{\mathrm{cc}}$ fixed and increasing $\sigma_{\mathrm{offset}}$, an effect that can be mimicked by a lower concentration in the NFW model. We run the risk of overfitting to the results. 

In fact, \citet{Johnston07} found very little constraining power on the miscentering width and the fraction of miscentered MaxBCG clusters, and applied strong priors to these distributions. \citet{George12} performed an extensive weak lensing miscentering study of groups in the Cosmological Evolution Survey (COSMOS), and chose to forgo the additional parameter $p_{\mathrm{cc}}$, as they achieved sufficiently good fits without it. \citet{Mandelbaum08} performed a lensing analysis of the MaxBCG clusters, and found that including miscentering effects with the \citet{Johnston07} prescription strongly affected the resultant fits for concentration, again asserting the degeneracies of these parameters. \citet{Mandelbaum08} conclude that this method of accounting for miscentering depends heavily on the mock catalogs from which the input parameters are generated, and in the case of MaxBCG clusters likely overcompensates. 

In a forthcoming paper, we will present weak lensing shear measurements of these clusters, as well as a more detailed investigation of the centroiding. Shear, being proportional to the differential surface mass density, is more affected by offset centers than magnification \citep{Johnston07}, and will be a better probe of miscentering.

\subsection{The Mass-Richness Relation}

\begin{figure*}
\begin{center}
\vspace{0.5 cm}
\includegraphics[scale=0.55]{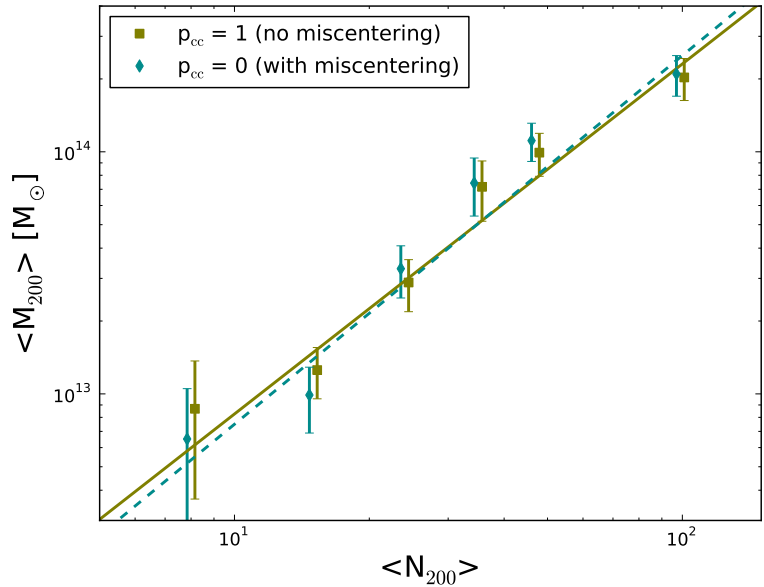}
\caption{Cluster mass-richness relation, using the same $N_{200}$ bins as in Figures \ref{binned}. Power law fits to the data are presented for both cases of with (blue diamonds and dashed line) and without (green squares and solid line) the effects of miscentering. Points are slightly offset horizontally for clarity.}
\vspace{1 cm}
\label{massrich}
\end{center}
\end{figure*}

We observe a prominent scaling of best fit mass to richness, across the six richness bins (although the first two bins do generally have overlapping error bars). We plot this trend in Figure \ref{massrich}, showing the average of the fit masses as a function of average cluster richness in each bin. Note that the distribution of $N_{200}$ in a bin is not uniform, and in the case of the highest richness bin the distribution is highly skewed (see Figure \ref{hist}).

We fit a simple power-law, Equation \ref{mr}, to these points, using the same plotted color and line schemes for perfectly centered and miscentered clusters. For this cluster sample, we find the best fit gives the normalization and slope of the mass-richness relation to be
\begin{equation}
M_0 = (2.3 \pm 0.2) \times 10^{13} M_\odot, \beta = 1.4 \pm 0.1
\end{equation}
for the perfectly centered $p_{\mathrm{cc}}=1$ case, and
\begin{equation}
M_0 = (2.2 \pm 0.2) \times 10^{13} M_\odot, \beta = 1.5 \pm 0.1
\end{equation}
for the miscentered $p_{\mathrm{cc}}=0$ case. The reduced $\chi^2$ are 0.9 and 1.7, respectively (4 degrees of freedom), and there is good agreement between the two different centering scenarios explored here.

It is difficult to directly compare the results for the mass-richness relation in this work to other studies. The main reason is that the richness $N_{200}$ we use is different than other definitions, which often count only red-sequence galaxies. Some uncertainty exists in the measure of richness as well, which we do not include in the analysis. Alternative measures of cluster richness would yield different scaling relations. Another factor is the cluster sample, which was compiled using a novel cluster-finder, and may well have different characteristics than other samples in the literature. In a follow-up paper we will present a shear analysis of the CFHTLenS clusters, and compare the mass-richness relation obtained using that complementary probe of halo mass.

\subsection{Redshift Binning}
\label{zbin}

\begin{figure*}
\begin{center}
\includegraphics[scale=0.8]{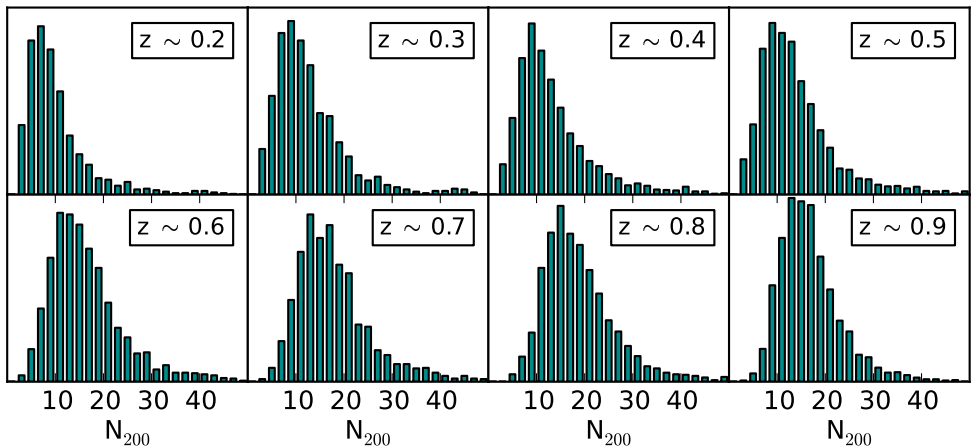}
\caption{$N_{200}$ distributions as a function of cluster redshift.}
\label{zbinhist}
\end{center}
\end{figure*}

\begin{figure*}
\begin{center}
\vspace{1 cm}
\includegraphics[scale=0.8]{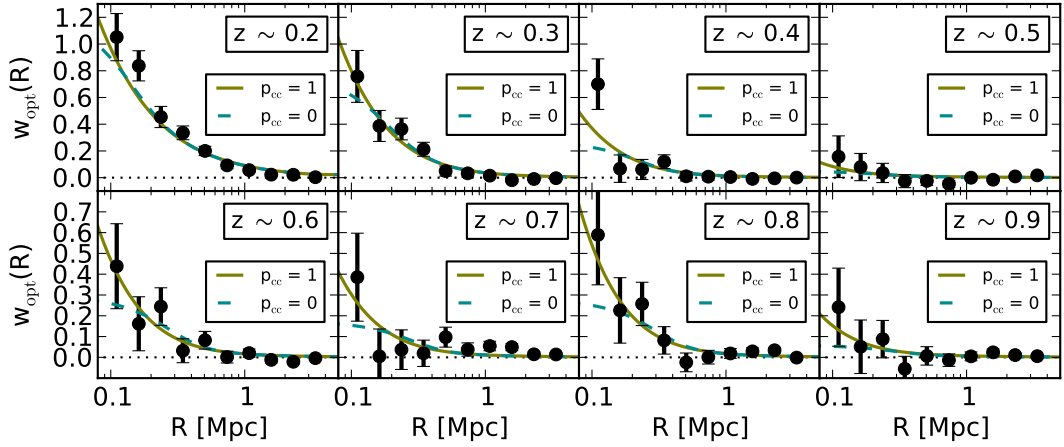}
\caption{Optimal correlation for clusters binned in redshift.}
\label{zbinw}
\end{center}
\end{figure*}

Finally, we investigate the magnification as a function of redshift. We stack clusters of all richnesses, at each redshift in the catalog, $0.2 \leq z \leq 0.9$, and measure the optimal correlations in each. This is displayed in Figure \ref{zbinw}. We observe a steady decrease in measured signal as the cluster redshift increases from $z \sim$0.2 to 0.5, then roughly consistent measurements from 0.6$\leq z \leq$0.8, followed by rather low signal at $z \sim$0.9. 

The $N_{200}$ distributions in Figure \ref{zbinhist} show that these trends cannot be caused by deviations in richness between these different cluster redshifts. This is difficult to reconcile with the clear mass-richness scaling observed when all redshifts are combined. Table 3 shows that detection significance for each reshift bin is more tightly linked to mass than the $\langle N_{200} \rangle$. Perhaps the richness estimates used in this work are not optimized for use as a mass proxy at all redshifts. Another possibility is that we have not correctly accounted for redshift overlap between samples. If the contamination fraction is higher than estimated, this could lead to a boost in correlation strength at low redshift, as well as a depletion at higher redshift. However it is still very difficult to explain the anomalously low measurement at intermediate redshift, $z \sim$0.5, with this reasoning.

One factor that we have not accounted for is possible cluster false detections in our sample. Since 3D-MF was optimized to produce cluster catalogs that are as complete as possible, false detection rates could be quite high. In particular, we would expect these rates to increase at high redshift, which would also weaken those measured correlations. We note in particular that cluster redshift bins $z \sim$ 0.5 and 0.9, which yield relatively low cluster masses, are seen in Figure \ref{pofz} to have excess numbers of detected clusters, possibly an indication of higher false detection rates at these redshifts.

\begin{table*}
    \label{ztable}
    \caption{Details of fits for redshift-binned measurements in Figure \ref{zbinw}. We list the same bin properties and fits given in Table \ref{richtable}, as well as $f_{clustering}$, which is the total fraction of LBGs expected to lie within $\Delta z \sim$0.1 of the cluster z.}
    \begin{tabular}{lllllllll}
      \hline
      Redshift & $f_{clustering}$ & Clusters & Significance & $\langle N_{200} \rangle$ & p$_{cc}$=1: $\langle M_{200} \rangle$ & $\chi^2_{red}$  & p$_{cc}$=0: $\langle M_{200} \rangle$ & $\chi^2_{red}$ \\ \hline
      $z$ $\sim$ 0.2 & 0.07 & 1157 & 12.5$\sigma$ & 11.6 & (9$\pm$2$\pm$2$^{sys}$)$\times10^{13} M_{\odot}$ & 3.6 & (9$\pm$2$\pm$2$^{sys}$)$\times10^{13} M_{\odot}$ & 3.4  \\
      $z$ $\sim$ 0.3 & 0.02 & 1515 & 8.0$\sigma$ & 14.4 & (6$\pm$1$\pm$1$^{sys}$)$\times10^{13} M_{\odot}$ & 2.2 & (6$\pm$1$\pm$1$^{sys}$)$\times10^{13} M_{\odot}$ & 2.1  \\
      $z$ $\sim$ 0.4 & 3$\times10^{-3}$ & 2242 & 4.6$\sigma$ & 15.2 & (1.9$\pm$0.7)$\times10^{13} M_{\odot}$ & 1.4 & (1.6$\pm$0.7)$\times10^{13} M_{\odot}$ & 1.6  \\
      $z$ $\sim$ 0.5 & 4$\times10^{-4}$ & 2932 & 4.0$\sigma$ & 15.9 & (0.3$\pm$0.4)$\times10^{13} M_{\odot}$ & 1.9 & (0.2$\pm$0.5)$\times10^{13} M_{\odot}$ & 1.9  \\
      $z$ $\sim$ 0.6 & 1$\times10^{-4}$ & 2455 & 4.6$\sigma$ & 18.0 & (2.2$\pm$0.8)$\times10^{13} M_{\odot}$ & 1.5 & (2.0$\pm$0.8)$\times10^{13} M_{\odot}$ & 1.6  \\
      $z$ $\sim$ 0.7 & 2$\times10^{-5}$ & 2331 & 4.5$\sigma$ & 19.3 & (1.2$\pm$0.7)$\times10^{13} M_{\odot}$ & 1.7 & (1.1$\pm$0.7)$\times10^{13} M_{\odot}$ & 1.9  \\
      $z$ $\sim$ 0.8 & 2$\times10^{-5}$ & 2364 & 4.9$\sigma$ & 19.9 & (2.5$\pm$0.9)$\times10^{13} M_{\odot}$ & 1.5 & (2.2$\pm$0.9)$\times10^{13} M_{\odot}$ & 1.7  \\ 
      $z$ $\sim$ 0.9 & 2$\times10^{-5}$ & 3040 & 2.6$\sigma$ & 17.6 & (0.5$\pm$0.5)$\times10^{13} M_{\odot}$ & 0.6 & (0.3$\pm$0.6)$\times10^{13} M_{\odot}$ & 0.8  \\
      \hline
    \end{tabular}
\end{table*}


\section{Conclusions}
\label{conc}
We present the most significant magnification-only cluster measurement to date, at 9.7$\sigma$. A sample of 18,036 cluster candidates has been detected using the 3D-MF technique in the $\sim$154 deg$^2$ CFHTLenS survey. In this analysis we have investigated the mass of cluster dark matter halos, from flux magnification, as a function of both richness and redshift. A forthcoming paper will present the weak lensing shear analysis of these clusters as well.

We fit a composite-NFW model that accounts for the full redshift and mass ranges of the cluster sample, as well as redshift overlap with low-$z$ source contaminants, cluster halo miscentering, and the 2-halo term. We find that the entire cluster sample is marginally better fit by the model that does not include miscentering, but do not see a strong preference either way across richness bins. In the future, shear measurements, which are more sensitive to miscentering, may illuminate this aspect of the investigation.

We observe a strong scaling between measured mass and cluster richness, and fit a simple power-law relation to the data. The two miscentering models explored in this work yield consistent values for the normalization and slope of the mass-richness relation.

We have attempted to account for the contamination of our background sources with low-$z$ galaxies. This is a serious systematic effect for magnification, as redshift overlap between lenses and sources will lead to physical clustering correlations, swamping the lensing-induced correlations that we want to measure. We use the full stacked redshift probability distributions for the lens and source populations, and include the expected clustering contribution in our model. In spite of this we see unexpected features in the redshift-binned measurements. Part of the reason could come from cluster false detections, which can be high for the 3D-MF method which is optimized for completeness. Another contribution could come from errors in the source redshift distributions. Accounting for redshift overlap is imperative if significant overlap exists between the lens and source distributions, or else mass estimates can end up very biased.

This is the first analysis presented of the 3D-MF clusters in CFHTLenS, but much more science is left to do with the sample. In particular, a more thorough investigation of the miscentering problem will be carried out in the forthcoming shear analysis, where it will be possible to compare different candidate centers. Another interesting question is whether dust can be detected on cluster scales by simulataneously measuring the chromatic extinction along with flux magnification. Finally different background source samples may be employed to improve signal-to-noise, but only if their redshift distributions can be well determined. We leave these tasks to future work.

This work has been an important step in the development of weak lensing magnification measurements, and the progression from signal detection to science. Many upcoming surveys will benefit from the inclusion of magnification in their lensing programs, as the technique offers a very complimentary probe of large scale structure. Since measuring flux magnification is not a strong function of image quality, it is especially useful for ground-based surveys which must deal with atmospheric effects. Next generation surveys like the Large Synoptic Survey Telescope (LSST), the Wide-Field Infrared Survey Telescope (WFIRST), and Euclid, will have greater numbers of sources, and improved redshift probability distribution estimates, so we can expect future magnification studies to yield important contributions to weak lensing science and cosmology.

\section*{Acknowledgements} 
We thank the CFHTLenS team for making their catalogues publicly available to download from www.cfhtlens.org. This work is partly based on observations obtained with MegaPrime/MegaCam, a joint project of CFHT and CEA/IRFU, at the Canada-France-Hawaii Telescope (CFHT) which is operated by the National Research Council (NRC) of Canada, the Institut National des Sciences de l’Univers of the Centre National de la Recherche Scientifique (CNRS) of France, and the University of Hawaii. This research used the facilities of the Canadian Astronomy Data Centre operated by the National Research Council of Canada with the support of the Canadian Space Agency. CFHTLenS data processing was made possible thanks to significant computing support from the NSERC Research Tools and Instruments grant program.

JF is supported by a UBC Four-Year-Fellowship and NSERC. HH is supported by the Marie Curie IOF 252760, a CITA National Fellowship and the DFG grant Hi 1495/2-1. LVW is supported by NSERC and CIfAR.

{\it Author Contributions:} JF led the analysis, wrote all code for the magnification measurement and modeling, and prepared the draft of this paper. HH and LVW contributed very significantly to the overarching approach and detailed form of the analysis, through countless meetings, emails, and discussions. In addition, HH is responsible for producing the photometric redshifts and LBG catalogs used in this work. TE did all the CFHTLenS data reduction. CL wrote the first version of the code used for calculating cluster richness. MM developed the 3D-MF cluster finding algorithm which was used to produce the entire CFHTLenS cluster candidate sample. CM performed an analysis of LBG contamination to determine the fraction of sources at low redshift.

\bibliography{References}

\end{document}